\documentclass[12pt]{article}
\usepackage[dvips]{graphicx}

\begin{document}

\begin{center}
{\Large\bf Electronic structure, hyperfine interactions and
disordering effects in iron nitride Fe$_{4}$N} \\

\vspace{0.5cm} {A.N. Timoshevskii,$^{\rm (1)}$ V.A.
Timoshevskii,$^{\rm (2)}$ B.Z. Yanchitsky$^{\rm (1)}$ and V.A.
Yavna $^{\rm (2)}$}\\

\vspace{0.6cm} {\it $^{\rm (1)}$ Institute of Magnetizm, 36-b
Vernadskii St., 252680 Kiev, Ukraine}
\\
{\it $^{\rm (2)}$ Rostov State University of Transport, Narodnogo
Opolcheniya 2, Rostov-on-Don, 344038 Russia}
\\
\end{center}

\begin{abstract}
Iron nitride Fe$_4$N is studied by full-potential LAPW method.
Structure para\-meters, electronic and magnetic properties as well
as hyperfine interaction parameters are obtained. We observe
perfect agreement with experimental results. Hypothetical Fe$_4$N
structure was also calculated to study the influence of
disordering effects on parameters of M\"ossbauer spectra. We
performed detailed analysis of EFG formation on Fe nuclei
including magnetization effects. We show that the formation of
N-Fe-N local configuration is energetically favourable in
nitrogen austenites.
\end{abstract}

\vspace{0.2cm}

\section{Introduction}

Face centered cubic (fcc) iron-based alloys are widely used for
developing of stainless austenitic steels especially for using in
critical temperature ranges, aggressive environ\-ment and other
severe external conditions. Doping of these steels with light
interstitial impurities (C,N) influence mechanics and kinetics of
structure phase transitions in Fe-based alloys. Distribution of
carbon and nitrogen atoms in solid solutions influence electrical
and mechanical properties of alloys. Nitrogen doping enables to
solve the problem of the strengthening of stainless steels.

Investigation of the influence of nitrogen on physical properties
of multicomponent systems is a complicated problem. The solution
of this problem should be made in several stages. On the first
stage it seems important to study electronic structure of iron
nitride Fe$_{4}$N, because binary $\gamma$-FeN$_x$ alloy can be
modeled by non-stoichiometric nitride $\gamma$-Fe$_{4}$N$_{1-x}$.

There are a lot of experimental data about atomic and electronic
structure of Fe$_{4}$N. We believe that M\"ossbauer spectroscopy
gives most interesting information about impurity distribution,
electronic structure and magnetic interactions in alloys.
Studying of hyperfine structure of the energy spectra of nuclei is
a powerful tool for investigation of interactions of atomic
nucleus with local electric and magnetic fields. These
interactions cause shifts and splittings of nuclear energy levels
and enable us to obtain information about symmetry of charge
distribution around the nucleus, about electronic configurations
of atoms and ions, as well as about peculiarities of atomic
structure of solids.

A number of experimental papers show substantial differences in
M\"ossbauer spectra of binary Fe-N and Fe-C alloys. These
differences are believed to be connected with different C and N
distribution in Fe fcc alloys \cite{gav1}. In this paper we
present calculation results of hyperfine interaction parameters
for iron nitride Fe$_{4}$N as well as for hypothetical Fe$_{4}$N
structure with another distribution of nitrogen atoms. This
allows us to determine changes in M\"ossbauer spectra caused by
redistribution on nitrogen atoms.

\section{Calculation details}

WIEN97 programme package \cite{wien}, employing Full-potential
Linearized Augmented Plane Wave (FLAPW) method was used for
calculations. As far as FLAPW is an all-electron method (unlike
pseudopotential methods), it allows to perform calculations of
hyperfine interaction parameters from first principles. Obtained
theoretical parameters of interaction of a nucleus with electric
and magnetic fields can be successfully compared with parameters
of experimental M\"ossbauer spectra.

Generalized gradient approximation (GGA) according to
Perdew-Burke-Ernzerhof \cite{pbe} model was used for
exchange-correlation potential. The radii of atomic spheres were
chosen as 1.9 a.u. and 1.68 a.u for Fe and N atoms respectively.
The accuracy of calculation results depends on several basic
parameters: number of $K$-points in Brillouin zone, number of
$LM$-components and Fourier coefficients in charge density and
potential decomposition and number of plane waves in interstitial
region. The choice of the values of these parameters was based on
convergence condition. Convergence tests gave the value
$R_{min}\times K_{max}=8.8$, which corresponds to 205 plane waves
per atom in the basis set. Inside atomic spheres the wave
function was decomposed up to $l_{max}=12$. Charge density and
potential was decomposed inside atomic spheres using lattice
harmonics basis up to $L_{max}=6$. In the interstitial region
Fourier expansion was used with 850 coefficients. Calculations
were performed for 3000 K-points in the Brillouin zone (84
K-points in the irreducible part). The values of all parameters
ensure accuracy of 0.1 mRy in total energy of the system. Due to
ferromagnetic nature of iron nitride all calculations were
performed using spin-polarized approximation.

\section{Atomic structure}

\begin{figure}[!htb]
\begin{center}
\includegraphics[scale=0.25,angle=-90]{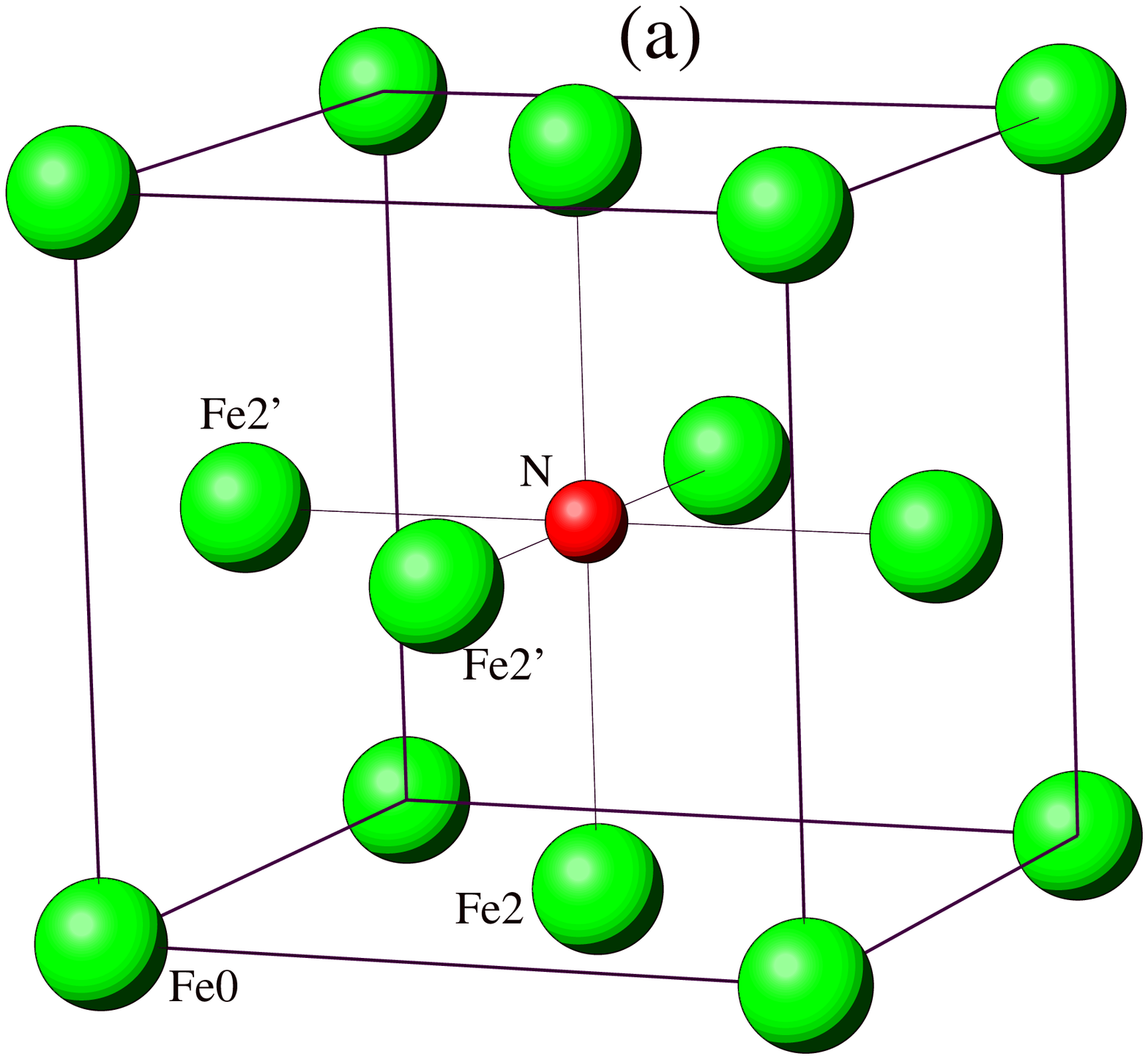}
\hspace{1.5cm}
\includegraphics[scale=0.3,angle=-90]{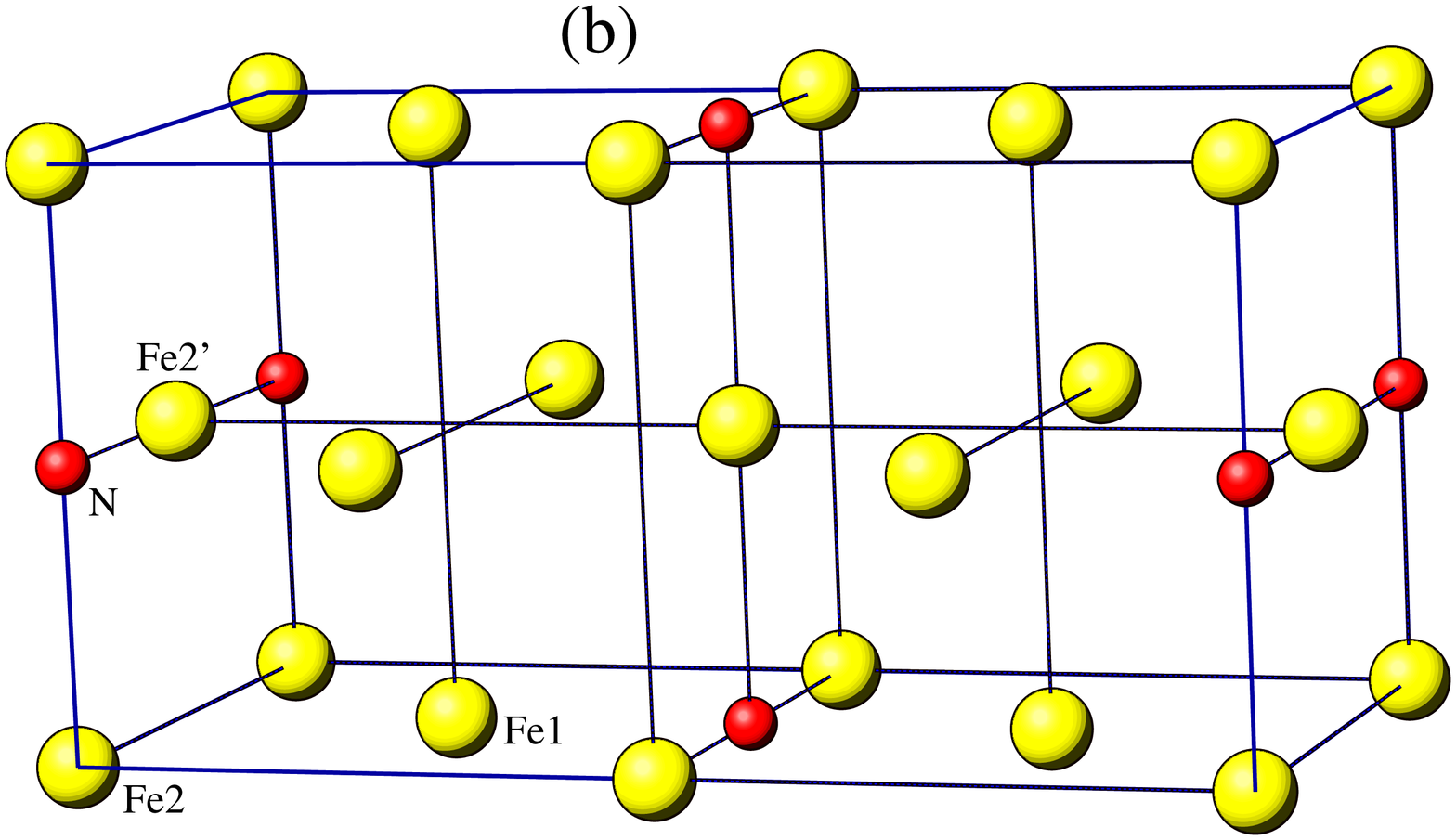}
\end{center}
\caption{The unit cell of iron nitride (a) and of hypothetical
Fe$_4$N structure (b)} \label{cell}
\end{figure}

The unit cell of iron nitride Fe$_4$N is a unit cell of
$\gamma$-Fe with a nitrogen atom introduced in the centre of the
cube (figure \ref{cell}a). The structure has two symmetry types of
Fe atoms: Fe$_2$ type forms octahedron around impurity atom and
has two impurity atoms in the first coordination sphere located
at $180^o$; Fe$_0$ type is located in the corners of the cell and
has no impurity atoms in the first coordination sphere.

In order to find the value of lattice parameter, corresponding to
the minimum total energy of the system we performed calculations
for five different values of lattice parameter. The results were
approximated by second-order polynomial using least square fit
method. Then the value of lattice parameter was obtained
analytically. Calculated value of lattice parameter C=7.164 a.u.
is in good agreement with experimental value C=7.17 a.u.
\cite{jacobs}. Calculations of electronic structure and hyperfine
interaction parameters were performed using obtained optimized
value of lattice parameter.

Optimization procedure was also performed for another two
systems: $\gamma$-Fe (Fe-fcc) and hypothetical Fe$_4$N structure
(figure \ref{cell}b). Both calculations were also performed in
spin-polarized approxi\-mation. Calculation of $\gamma$-Fe seems
interesting because it allows us to determine the influence of
introducing of nitrogen atom on electronic structure and
hyperfine interaction parameters of Fe-fcc. Optimization of
hypothetical Fe$_4$N structure (figure \ref{cell}b) was performed
including variation of interatomic Fe-N distance. This structure
has two types of iron atoms: Fe$_1$ and Fe$_2$, having one and
two N atoms in the nearest neighbourhood respectively. We found
out that with this type of ordering the total energy of the
structure is 18 mRy larger, than the total energy of iron
nitride. This shows that the formation of Fe-N-Fe-N chains
(Fe$_2$ configuration) without Fe-N pairs (Fe$_1$ configuration)
is energetically favourable.

From the point of view of atomic structure of Fe-N alloys iron
nitride Fe$_{4}$N is the simplest structure where Fe-N chains are
combined with Fe-Fe chains. This makes possible to perform
detailed analysis of the influence of nitrogen on electronic
structure and hyperfine interaction parameters and to use
obtained information later for analyzing more complicated Fe-N
alloys. This step-by-step approach, based on application of
up-to-date \textit{ab initio} calculation method, appears to be
most accurate for solving the problem of the influence of
impurity atoms on the properties of Fe-based alloys.

\begin{table}[!htb]
\begin{footnotesize}
\caption{\footnotesize{Cell parameters (a,b,c) and magnetic
moments (M) for $\gamma$-Fe, iron nitride and model Fe$_4$N
structure. Cell parameter is presented in atomic units, magnetic
moment - in Bohr magnetons.}} \label{tab1}
\begin{itemize}
\item[]\begin{tabular}{p{3cm} p{2cm} p{2cm} p{2cm} p{2cm}}
 \hline
 \hline
 Structure & $a,b,c$ (a.u)& $a,b,c$ (a.u.)  & M($\mu_{B}$) &M($\mu_{B}$)\\
           & theory     & expt.    & theory       & expt.\\
 \hline
 \hline
 Fe(fcc)   & $a=6.58$       & $a=6.89^a$  &  $1.04$        &        \\
 \hline
 Fe$_{4}$N & $a=7.164$      & $a=7.171^b$ &              &          \\
 Fe$_0$    &              &           &  2.90        & $3.0^c$    \\
 Fe$_2$    &              &           &  2.31        & $2.0^c$    \\
 Aver./atom  &            &           &  2.50        & $2.21^c$  \\
 \hline
 Fe$_{4}$N (model) & a=7.114      &        &              &          \\
           & $c=14.266$      &        &              &          \\
 Fe$_1$    &            &           &  2.53        &     \\
 Fe$_2$    &            &           &  1.98        &     \\
 Aver./atom  &          &           &  2.27        &   \\
 \hline
 \hline
\end{tabular}
\item[]\cite{acet}$^a$, \cite{jacobs}$^b$, \cite{coeh}$^c$
\end{itemize}
\end{footnotesize}
\end{table}

\section{Hyperfine interactions}

Ferromagnetic nature of iron nitride Fe$_{4}$N makes M\"ossbauer
spectrum of this compound much more complicated. The presence of
magnetic field leads to combined magnetic and electric hyperfine
interactions. The Hamiltonian of the system, including these
interactions is given by
\begin{equation} \label{ham1}
\hat{\mathcal{H}}=-(\vec{\mu}\times\vec{H})+
e\sum_{ij}Q_{ij}V_{ij},
\end{equation}
where $\vec{\mu}$ - magnetic moment of the nucleus, $\vec{H}$ -
magnetic field, $Q_{ij}$ - tensor of the nuclear quadrupole
moment, $V_{ij}$ - tensor of electric field gradient (EFG). The
principal EFG axis and magnetization axis may be non-collinear,
which makes finding of Hamiltonian eigenvalues much more
complicated. Using $eQ/ \mu H \ll 1$ approximation and taking
into account that $I=3/2$ for excited state of $^{57}$Fe nucleus,
the energy levels of the system are given by
\begin{eqnarray} \label{ham3}
E = -g\mu_{n}H m_I + (-1)^{|m_I|+1/2} \frac{eQV_{zz}}{4}
\frac{3\cos^2\theta-1}{2}\left(1+\frac{\eta^2}{3}\right)^{1/2}
\\ m_I=-I,-I+1,\ldots,I-1,I \nonumber
\end{eqnarray}
In this expression $m_I$ - magnetic quantum number, $V_{zz}$ -
the principal component of EFG tensor, $\eta$ - asymmetry
parameter, $\theta$ - angle between the principal axis of EFG
tensor and the direction of magnetic field $\vec{H}$.

The presence of magnetic field reduces the symmetry of the
system. In iron nitride Fe atoms, located in the centre of the
cube faces are now divided into two types: Fe$_2$, for which
principal EFG axis is parallel to magnetic field, and Fe$_2'$,
for which this axis is perpendicular to magnetic field (figure
\ref{cell}a). M\"ossbauer spectra of these two types of Fe atoms
will have different quadrupole splitting. So, due to magnetic
field, we have three different types of iron atoms in iron
nitride structure. The total M\"ossbauer spectrum of this
compound will consist of eighteen peaks. Twelve of these peaks,
which belong to Fe$_2$ and Fe$_2'$ atoms, will have different
shifts due to different quadrupole interactions. The ratio of the
number of these atomic types in iron nitride unit cell is
Fe$_0$:Fe$_2$:Fe$_2'$=1:1:2, which leads to the same ratio of
intensities of respective M\"ossbauer peaks. All described
features of the spectrum are totally approved by experimental
data \cite{sp1}.

Quadrupole splitting, expressed in $mm/s$, is given by
\begin{equation}\label{quadro1}
\Delta_v [mm/s]=0.5205 \times Q[b] V_{zz}[10^{21}v/m^2]
\frac{3\cos^2\theta-1}{2}\left(1+\frac{\eta^2}{3}\right)^{1/2},
\end{equation}
In our calculations we used the value of nuclear quadrupole moment
Q$^{57}$Fe =0.16$b$. This value was determined by Dufek, Blaha and
Schwarz \cite{fe57} by comparing experimental quadrupole
splitting and theoretical EFG values, calculated by FLAPW method,
for fourteen different Fe compounds.

For temperatures above the Curie point ($T_c$) the system becomes
paramagnetic, and Fe$_0$ atoms will not contribute to M\"ossbauer
spectrum. Fe$_2$ and Fe$_2'$ atomic types become totally
identical, and M\"ossbauer spectrum in this case is a doublet
with quadrupole splitting
\begin{equation}\label{quadro2}
\Delta_v [mm/s] = 1.041 \times Q[b] V_{zz}[10^{21}v/m^2]
\left(1+\frac{\eta^2}{3}\right)^{1/2}.
\end{equation}

In this paper electric field gradient was calculated according to
method, developed by Blaha \textit{et al} \cite{efg1}. According
to this approach EFG is calculated on \textit{ab-initio} basis
directly from electronic density distribution. We obtain EFG
contributions from charge density inside atomic spheres
(\textit{valence} EFG) and outside the spheres (\textit{lattice}
EFG). Valence EFG is calculated as an integral of the value
$\rho_{2M}(r)/r$ over the atomic sphere \cite{efg2}. The values
$\rho_{LM}$ originate from two radial wave functions with $l$ and
$l'$:
\begin{equation}\label{rho}
\rho_{LM}(r)=\sum_{E<E_F}\sum_{l,m}\sum_{l',m'} R_{lm}(r)
R_{l'm'}(r) G_{Lll'}^{Mmm'},
\end{equation}
where $R_{lm}(r)$ - LAPW radial wave functions, $G_{Lll'}^{Mmm'}$
- Gaunt integrals.

Table \ref{tab2} presents calculated principal values of EFG
tensor and obtained according to (\ref{quadro1}) values of
quadrupole splitting (QS) for Fe$_2$ and Fe$_2'$ atoms. The table
shows that our results are in perfect agreement with experimental
data, obtained by Foct \textit{et al} \cite{sp1} for the sample
at T=4K. Excellent agreement with experiment also shows that
chosen direction of magnetic axis (001) is a direction of real
magnetic field in iron nitride Fe$_{4}$N.

Using (\ref{quadro2}) we also calculated quadrupole splitting at
$T>T_c$. Our value $\Delta_v$=0.50 mm/s is also in excellent
agreement with experimental data of Foct (0.50 mm/s), obtained at
T=763K \cite{sp1}.

\begin{table}[!tb]
\begin{footnotesize}
\caption{\footnotesize{Contributions to $V_{zz}(10^{21}v/m^2)$
from inside the spheres and from interstitial region. Theoretical
and experimental values of quadrupole splitting (QS)(mm/s) for Fe
atoms with different angle $\theta$ between $\vec{H}$ and
principal EFG axis.}} \label{tab2}
\begin{tabular}{p{1.0cm} p{0.7cm} p{1.5cm} p{1.5cm} p{1.5cm} p{0.7cm} p{0.7cm} p{1.5cm} p{2.0cm}}
 \hline
 \hline
 System & Atom   & Valence & Lattice & Total  &$\eta$ &$\theta ^o$&QS     & QS  \\
        &        & EFG     & EFG     & EFG    &       &           & theory& expt. \cite{sp1}\\
 \hline
 \hline
 Nitride \\
         & Fe$_2$ & $-3.369$  & $+0.382$   & $-2.987$ & 0   &$0^o$  & $-0.25$  & $-0.26\pm 0.02$\\
         & Fe$_2'$& $-3.369$  & $+0.382$   & $-2.987$ & 0   &$90^o$ & $\;\;\:0.12$&$\;\;\:0.12\pm0.02$\\
 Model \\
        & Fe$_1$&   $-0.34$  & $+0.14$     & $-0.20$  & 0   &$90^o$ &$\;\;\:0.01$     & \\
        & Fe$_2$ &  $-2.41$  & $+0.62$     & $-1.79$  &0.76 &$0^o$  & $-0.16$    & \\
        & Fe$_2'$ &  $-2.41$  & $+0.62$     & $-1.79$  &0.76 &$90^o$&$\;\;\:0.08$ & \\
 \hline
 \hline
\end{tabular}
\end{footnotesize}
\end{table}

Table \ref{tab2} also presents calculation results for model
Fe$_4$N structure. Magnetization axis is assumed to be directed
along one of Fe-N-Fe-N chains. Redistribution of nitrogen atoms
leads to substantial changes of EFG values. The quadrupole
splitting of Fe$_2$ atoms is reduced by 1/3 of the previous value
when Fe$_1$ configuration appears in the structure. The same
effect we observe at T$>$T$_c$: quadrupole splitting becomes now
0.33 mm/s. We see that quadrupole splitting is very sensitive to
redistribution of impurity atoms in Fe-fcc. This fact can be used
for interpretation of M\"ossbauer spectra of real fcc-alloys with
different distribution of interstitial impurity.

\section{Charge density and electric field gradient}

We performed detailed analysis of EFG formation in iron nitride
Fe$_{4}$N. Table \ref{tab2} shows that valence component makes
90\% contribution to the total value. So, we can focus on this
component for understanding of EFG origin.

According to our approach the valent band of the compound
(Fe-$3s3p3d$, N-$2s2p$) was divided to four energy regions
according to table \ref{tab3}. These energy intervals are formed
mostly by different electronic shells of iron and nitrogen atoms.
We also analyzed EFG contributions from wavefunctions of different
symmetry (according to (\ref{rho})) for each of energy regions.

\begin{table}[!htb]
\begin{footnotesize}
\caption{\footnotesize{Principal EFG component (10$^{21}$
$v/m^2$) for iron nitride Fe$_{4}$N. Contributions to V$_{zz}$
from different energy intervals and from radial wavefunctions of
different symmetry.}} \label{tab3}
\begin{tabular}{p{2.0cm} p{2.0cm} p{2.0cm} p{2.0cm} p{2.0cm}}
 \hline
 \hline
 Energy   & $s-d$   & $p-p$   & $d-d$    & Total \\
 region   &         &         &          &        \\
 \hline
 \hline
 Fe-$3s3p$ & $-0.046$  &$ +4.437$  &$ -0.001$   & $+4.390$  \\
 N-$2s$    & $-0.021$  & $-7.889$  & $-0.364$   &$ -8.274$  \\
 N-$2p$    & $-0.037$  & $-14.652$ & $-5.680$   & $-20.369$ \\
 Fe-$3d$   & $+0.092$  & $+7.892$  & $+13.121$  & $+21.105$ \\
 \hline
 Valent    & $-0.012$  & $-10.212$ & $+7.076$   & $-3.148$  \\
 \hline
 \hline
\end{tabular}
\end{footnotesize}
\end{table}

\begin{table}[!htb]
\begin{footnotesize}
\caption{\footnotesize{V$_{zz}$ contributions from N-$2p$ and
Fe-$3d$ energy regions for electrons with different spin
orientation.}} \label{tab4}
\begin{tabular}{p{1.5cm} p{1.5cm} p{1.5cm} p{1.5cm} p{1.5cm} p{1.5cm}}
 \hline
 \hline
 Energy & Spin    & $s-d$   & $p-p$    & $d-d$   & Total \\
 region &         &         &          &         &        \\
 \hline
 \hline
 N-$2p$ & up      & $-0.015$   & $-6.535$  & $-3.021$  & $-9.571$  \\
        & down    & $-0.022$   & $-8.117$  & $-2.659$  & $-10.798$ \\
 \hline
 Fe-$3d$& up      & $+0.042$    & $+2.216$ & $+10.190$ & $+12.448$  \\
        & down    & $+0.050$    & $+5.676$ & $+2.931$  & $+8.657$   \\
 \hline
 \hline
\end{tabular}
\end{footnotesize}
\end{table}

First of all let us analyze the electric field gradient, caused
by the whole valent band (valence EFG). Table \ref{tab3} shows
that $s-d$ contribution is negligibly small for all energy
regions. EFG is practically formed by $p-p$ and $d-d$
contributions, which have different signs. We see that negative
$p-p$ contribution is dominating, although it is greatly canceled
by positive $d-d$ contribution.

\begin{figure}[!b]
\begin{center}
\includegraphics[scale=0.35]{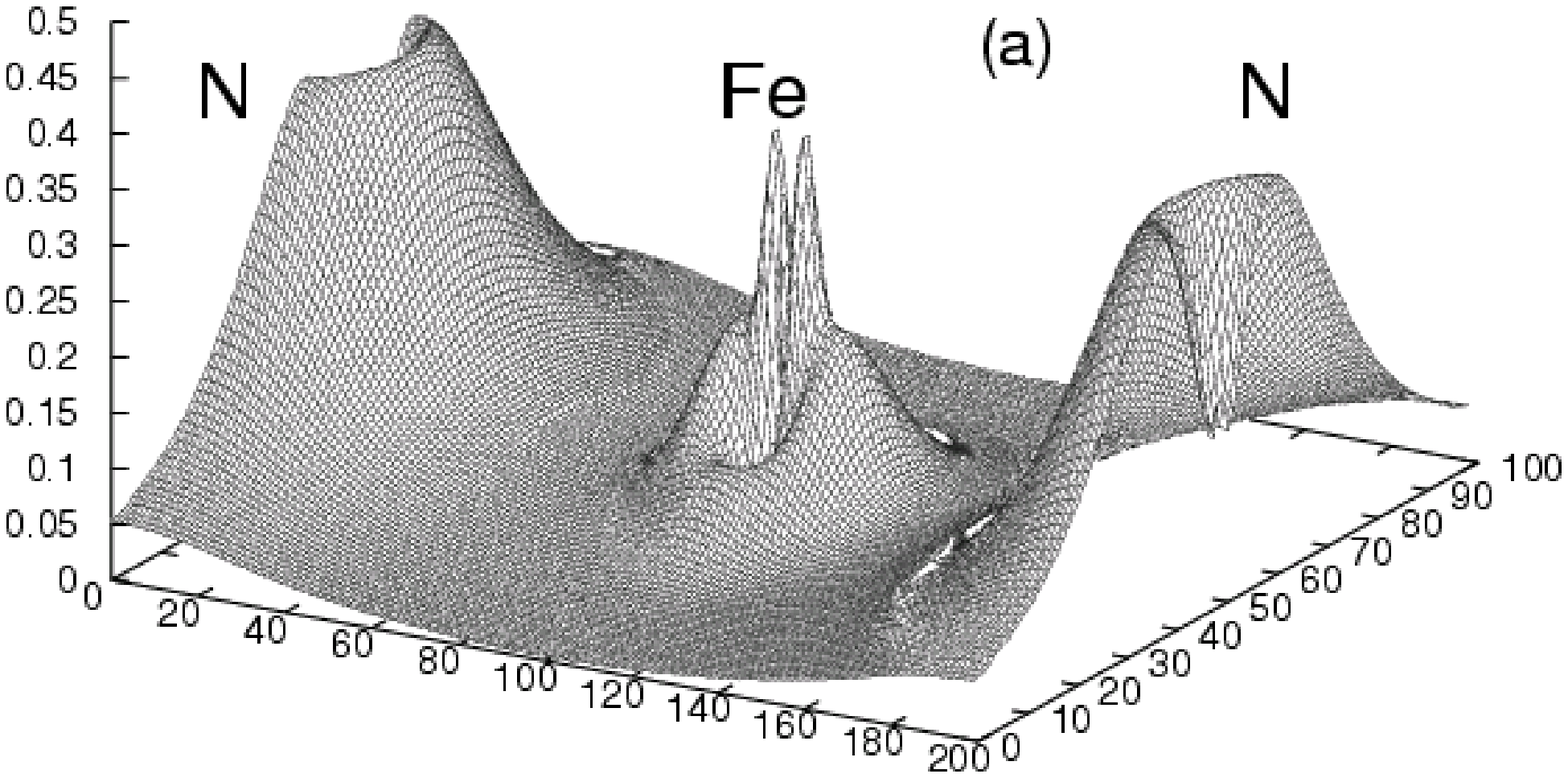}
\includegraphics[scale=0.35]{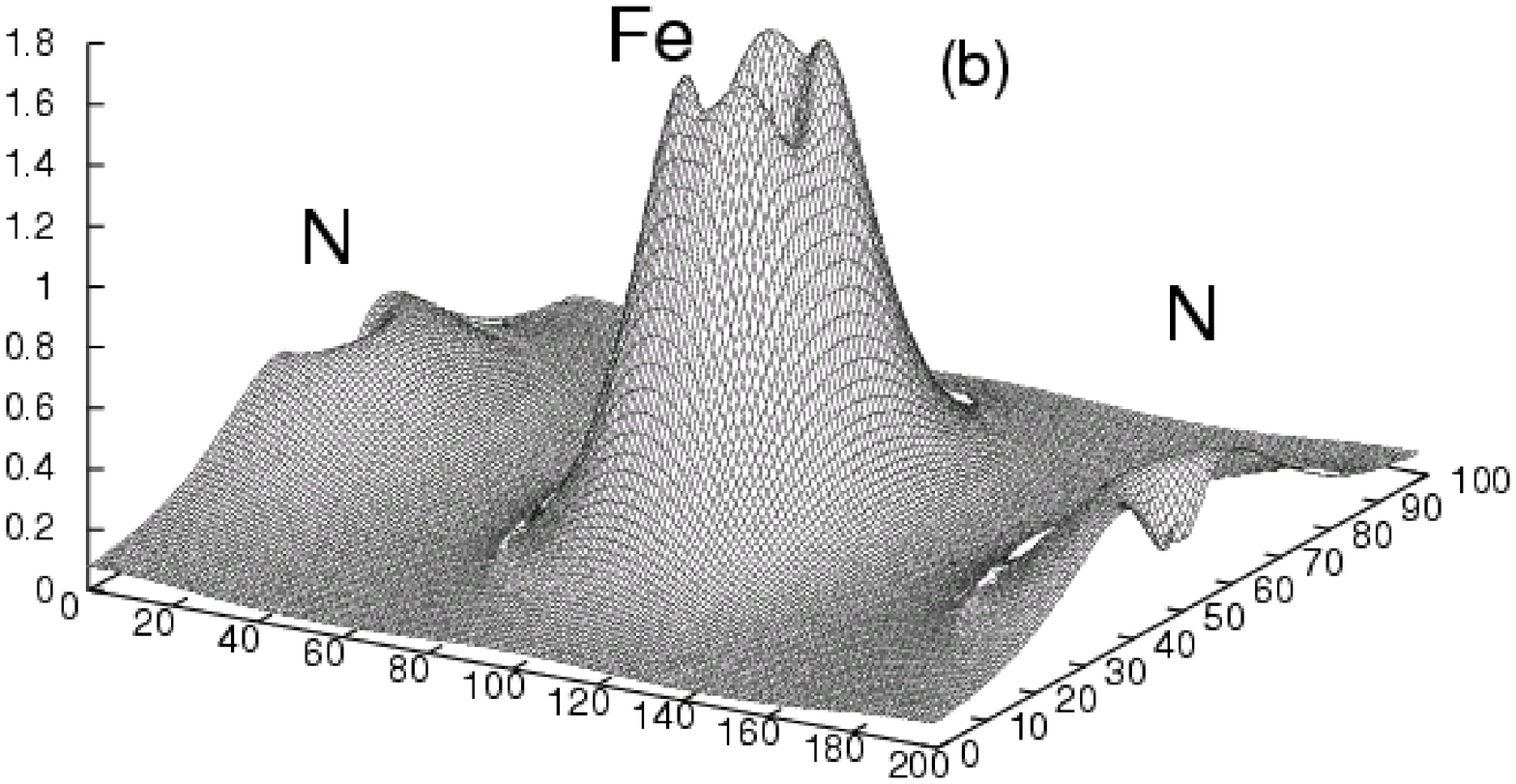}
\end{center}
\caption{\small{Charge distribution in N-$2p$ (a) and N-$2p$Fe$3d$
(b) energy region for iron nitride Fe$_4$N, $e/(a.u.)^3$}}
\label{den1}
\end{figure}

The analysis of table \ref{tab3} data allows us to conclude that
main contribution to EFG formation is made by $2p$ and $2s$
nitrogen states, and contribution from $2p$ states is dominant.
It should be noted that we are now talking about electrons, which
have energy of nitrogen $2p$ states, but spatially are located in
Fe$_2$ sphere. Substantial asymmetry of spatial distribution of
these electrons (figure \ref{den1}a) gives main negative
contribution to total EFG. The interesting fact is that more than
70\% of charge, formed by these electrons, have $p-$symmetry, and
30\% - $d-$symmetry relative to the centre of Fe$_2$ sphere. This
follows from analysis of $p-p$ and $d-d$ contributions for N-$2p$
interval (table \ref{tab3}). The same situation is observed for
$2s-$electrons of nitrogen, but their total EFG contribution is
2.5 times smaller than $2p-$electrons contribution. The EFG
contribution of N-$2s-$electrons is practically caused only by
electrons of $p-$symmetry relative to the centre of Fe$_2$ sphere.

Substantial negative EFG, caused by described asymmetry of
nitrogen electrons, is greatly canceled by positive gradient,
caused by own Fe electrons (figure \ref{den1}b). These are first
of all Fe-$3d-$electrons, which density is distributed in such a
way, that greatly cancels asymmetry of $2s$ and $2p$ electrons of
nitrogen. Comparative analysis of $p-p$ and $d-d$ contributions
shows that more than 60\% of asymmetrically distributed electrons
of Fe-$3d$ energy region have $d-$symmetry, and less than 40\% -
$p-$symmetry (table \ref{tab3}) relative to the centre of Fe$_2$
sphere. Redistribution of rather localized Fe-$3s3p$ states also
gives positive EFG contribution, but this contribution is almost
5 times smaller than contribution of $3d-$states.

It should be noted that quadrupole splitting was calculated using
total EFG value, given in table \ref{tab2}, and for calculation
of valence EFG contribution summation in (\ref{rho}) was done up
to $l_{max}=12$. On the other hand, for analysis of partial EFG
contributions we limited the summation only up to $l_{max}=2$
(due to the absence of "chemical" $f$-states). This is the reason
for small differences of valence EFG values in tables \ref{tab2}
and \ref{tab3}.

The values of local magnetic moments play an important role in
EFG formation in iron nitride Fe$_{4}$N. The calculations show
that the upper edge of Fe-$3d$ spin down states is located 2 eV
above the Fermi level. Substantial difference in population of
states with different spin orientation greatly influence EFG
formation. Table \ref{tab3} gives EFG contributions from states
with different spin orientation for N-$2p$ and Fe-$3d$ energy
regions. We notice considerable difference in EFG values, caused
by spin up and down electrons of Fe-$3d$ interval. This effect
mostly influence the value of $d-d$ contribution of this energy
region. This contribution is mainly formed by density
distribution of own $3d-$electrons of iron. Spin up contribution
is more than 3 times larger than spin down contribution for $d-d$
component of Fe-$3d-$region.

The fact that spin down states are partially "pulled" above the
Fermi level considerably decreases compensating positive
component, and leads to increasing of absolute value of total
(negative) EFG. In the absence of this effect we would have
considerably larger compensating component, which would lead to
decreasing of total EFG value and, as a consequence, to
decreasing of quadrupole splitting in M\"ossbauer spectra.

\section{Summary}
We performed \textit{ab initio} band structure calculations of
iron nitride Fe$_4$N. All our results are in very good agreement
with experimental data. Structure characteristics as well as
magnetic properties are well reproduced. We also performed
calculations of hypothetical Fe$_4$N disordered structure.
Comparative analysis of these two calculations showed that the
formation of Fe-N pairs in Fe$_4$N compound is energetically
unfavourable.

We also calculated hyperfine interaction parameters for both
Fe$_4$N structures taking into account the presence of magnetic
field. For iron nitride structure we obtained perfect agreement
of quadrupole splitting values with experimental data. For
disordered Fe$_4$N structure we observe considerable decreasing
of quadrupole splitting for Fe-N-Fe-N chains, and very small
quadrupole splitting for Fe-N pairs.

Detailed analysis of EFG formation in iron nitride Fe$_4$N allows
us to conclude that EFG is mainly formed by space distribution of
nitrogen $2s$ and $2p$ electrons, although it is greatly canceled
by Fe$3d-$electrons. Magnetization reduces this cancellation
effect and increases the value of total EFG in iron nitride
structure.


\begin{thebibliography}{99}
\bibitem{gav1} V.G.Gavriljuk \textit{et al}, Acta mater 45, 225
(1997), Acta mater 47, 927 (1999)
\bibitem{wien} P.Blaha, K.Schwarz, and J.Luitz, WIEN97, A Full Potential
Linearized Augmented Plane Wave Package for Calculating Crystal
Properties (Karlheinz Schwarz, Techn. Universit\"at Wien,
Austria), 1999. ISBN 3-9501031-0-4
\bibitem{pbe} Perdew J.P.,Burke S. and Ernzerhof M. 1996,Phys.Rev.Let.77,3865
\bibitem{fcc} J.H\"aglund, Phys. Rev. B 47, 566 (1993)
\bibitem{lo} D.Singh, Phys. Rev. B 43, 6388 (1994)
\bibitem{fe57} P.Dufek, P.Blaha and K.Schwarz, Phys. Rev. Lett.
75, 3545 (1995)
\bibitem{acet} M.Acet, H.Z\"ahres, E.F.Wasserman and W.Pepperhoff, Phys. Rev. B 49, 6012
(1994)
\bibitem{jacobs} H.Jacobs, D.Rechenbach and U.Zachwieja, Journal of Alloys and
Compounds 227, 10 (1995)
\bibitem{iron} H.C.Herper, E.Hoffmann and P.Entel, Phys. Rev. B 60, 3839 (1999)
\bibitem{efg1} P.Blaha, K.Schwarz and P.Herzig, Phys. Rev. Lett., 54, 1192 (1985)
\bibitem{efg2} K.Schwarz, C.Ambrosch-Draxl and P.Blaha, Phys. Rev. B, 42, 2051 (1990)
\bibitem{sp1} P.Rochegude and J.Foct, Phys. Stat. Sol. (A) 98, 51 (1986)
\bibitem{coeh} R.Coehoorn, G.H.O.Daalderop and H.J.F.Jansen, Phys. Rev. B, 48, 3830 (1993)

\end{thebibliography}
\end{document}